\documentclass[epjCONF,columns]{svjour} 
\session-title{2011 Hadron Collider Physics Symposium (HCP-2011)}
\usepackage{graphics}
\usepackage[varg]{txfonts} 
\usepackage[latin1]{inputenc}
\def\LowPt{{\mathsf{LowPt}}}
\def\HighPt{{\mathsf{HighPt}}}
\def\veryHighPt{{\mathsf{veryHighPt}}}
\def\met{\ensuremath{E_{\mathrm{T}}^{\mathrm{miss}}}} 

\def\Znn{\ensuremath{Z \  (\rightarrow \nu \bar{\nu} )}}
\def\Wtauj{\ensuremath{W (\rightarrow \tau\nu)}}
\def\Wenj{\ensuremath{W (\rightarrow e\nu)}}
\def\Wmnj{\ensuremath{W (\rightarrow \mu\nu)}}
\def\Ztt{\ensuremath{Z/\gamma^* (\rightarrow \tau^+ \tau^- )}}
\def\Zmm{\ensuremath{Z/\gamma^* (\rightarrow \mu^+ \mu^-)}}
\def\Zee{\ensuremath{Z/\gamma^* (\rightarrow e^+ e^-)}}
\def\Zll{\ensuremath{Z/\gamma^* (\rightarrow \ell^+ \ell^-)}}
\newcommand{\ptjet}{p_{\rm  T}}
\newcommand{\pt}{p_{\rm  T}}

\newcommand{\etajet}{\eta^{\rm jet}}

\session-title{2011 Hadron Collider Physics Symposium (HCP-2011)}
\begin{document}
\title{Search for new phenomena in events with a monojet and large missing transverse momentum at the LHC using the ATLAS detector}
\author{Mario Mart\'\i nez\inst{1}\fnmsep\thanks{\email{mmp@ifae.es}}}
\institute{ICREA/Institut de F\'\i sica d'Altes Energies, Bellaterra, Barcelona 08193, Spain. \\ \\
({\it{on behalf of the ATLAS Collaboration}})}
\abstract{
We report preliminary results on a search for new phenomena in an event sample with monojets and 
large missing transverse momentum in the final state. The analysis uses 1.00~fb${}^{-1}$ of 
data collected in 2011 with the ATLAS detector~\cite{atlas_eps}. Good agreement is observed between the 
number of events in data and the Standard Model predictions. 
The results are translated into improved limits on a model with Large Extra Dimensions.
} 
\maketitle
\section{Introduction}
\label{intro}

The search for new physics in events with a high transverse energy jet and large missing 
transverse energy constitutes one the simplest and most striking signatures that can be observed at a hadron 
collider. Different theoretical models for physics beyond the standard model (SM) predict the 
presence of {\it{monojet}} signatures in the final state like, for example,  
Large Extra Dimension (LED) scenarios~\cite{add}. 

In the Arkani-Hamed, Dimopoulos, 
and Dvali (ADD) LED model considered in this analysis,
the four-dimensional Planck scale, $M_{Pl}$, is related to the 
fundamental $4+n$-dimensional Planck scale, $M_D$, by ${M_{Pl}}^2 \sim {M_D}^{2+n}R^n$, where 
$n$ and $R$ are the number and compactification radius of the extra dimensions, respectively. 
An appropriate 
choice of $R$ for a given $n$ allows for a value of $M_D$ close to 
the electroweak scale. The compactification of the extra spatial dimensions results in 
a Kaluza-Klein tower of massive graviton modes.
These graviton modes are produced in association with a jet and do not interact with the detector, which results
in a monojet signature in the final state. For the production of the graviton mass modes 
in the ADD scenario, a low-energy 
effective field theory, governed by the energy scale $M_D$, is used.


\section{Event Selection}
\label{sec:event}

The data sample considered in this analysis was collected with
ATLAS tracking detectors, calorimeters, muon chambers, and magnets 
operational, and corresponds to a total integrated luminosity of 1.00~fb${}^{-1}$. 
The data were selected online using 
a trigger logic that selects events with missing transverse momentum $\met$ above 60~GeV, 
which is  more than 98$\%$ efficient for  
$\met > 120$~GeV, as determined using an unbiased data sample with muons in the final state. 

The offline event selection criteria applied follow closely those in Ref.~\cite{plb}, which defined two 
separate $\LowPt$ and $\HighPt$ set of requirements with the aim to maintain the sensitivity to a varity of 
models for new phenomena. With the increase in statistics, a new $\veryHighPt$ set of requirements 
is now defined to improve  the sensitivity to signals of new phenomena at very large transverse momenta.  
The following event selection criteria are applied:

\begin{itemize}


\item Events are required to have a reconstructed primary vertex. This 
rejects beam-related backgrounds and cosmic rays. 


\item Events are rejected if they contain any jet with $\ptjet > 20$~GeV and 
$|\eta| < 4.5$ that present anomalous charged fraction $f_{\small{\textrm{ch}}}$~\cite{atlas_eps}, 
electromagnetic fraction $f_{\small{\textrm{em}}}$ in the calorimeter, or timing (as determined from the 
energy deposits of the jet constituents) inconsistent with originating  
from a proton-proton collision, and most likely produced by beam-related backgrounds and cosmic rays.   
 In addition, the highest 
$p_{\rm T}$ jet selected (see below) is required to have $f_{\small{\textrm{ch}}} > 0.02$ and 
$f_{\small{\textrm{em}}} > 0.1$.
Additional requirements are applied to 
suppress coherent noise and electronic noise bursts in the calorimeter producing
anomalous energy deposits.

\item During 2011, part of the data suffered from the presence of a hole in the 
electromagnetic calorimeter coverage in the region $0 < \eta < 1.45$ and $0.788 < \phi < -0.592 $, which affected the readout of two of the LAr calorimeter layers.  The effect of the reduced calorimeter response was studied and 
resulted into a slightly modified event selection. 
A fiducial requirement is applied to the jets and electrons in the final 
state to avoid any bias in the analysis. Events are rejected if there is any jet with $\ptjet $ above 20~GeV or 
an identified electron in the final state such that their distance to the edges of the calorimeter affected 
region in $\eta - \phi$ is less than 0.4 or 0.1, respectively.

\item   Events are required to have no identified electrons or muons in the final state.

\item As in~\cite{plb}, the $\LowPt$ ($\HighPt$) selection requires a jet with $\ptjet > 120$~GeV 
($\ptjet > 250$~GeV) and $|\etajet| < 2$ in  the final state, and $\met > 120$~GeV ($\met > 220$~GeV). 
Events with a second leading jet $\ptjet$ above 30~GeV (60~GeV) in the region $|\eta| < 4.5$ are rejected. 
For the $\HighPt$ selection, the $\ptjet$ of the third leading jet must be less than 30~GeV, and an 
additional requirement on the azimuthal separation  $\Delta\phi(\textrm{jet},\met) > 0.5$ 
between the $\met$ and the direction of the second leading jet is required, that reduces 
the QCD background contribution where the large $\met$ originates from the mis-measurement 
of the second-leading jet $\ptjet$. 

\item A new $\veryHighPt$ selection is defined with  the same requirements as in the $\HighPt$ region, but with 
thresholds on the leading jet $\ptjet$ and $\met$ increased up to 350~GeV and 300~GeV, respectively.

\end{itemize}

\noindent
A total of 15740, 965 and 167 events are observed in the $\LowPt$, $\HighPt$ and $\veryHighPt$ regions, respectively.

\section{Background estimation}
\label{sec:backg}

The expected background to the monojet signature is dominated by $\Znn$+jets and   
$W$+jets production, and  includes contributions from $\Zll$+jets 
($\ell = e, \mu, \tau$), multi-jets, $t\bar{t}$, 
and $\gamma$+jets processes. The W/Z plus jets backgrounds are estimated using 
Monte Carlo  (MC) event samples normalized using data in control regions.
The remaining SM backgrounds from $t\bar{t}$ and $\gamma$+jets 
are determined using simulated samples, while the multi-jets background  contribution
is extracted from data. Finally, the potential
contributions from beam-related background and cosmics rays are
estimated using data.  

\subsection{Electroweak background}

As explained in~\cite{plb}, control samples in data, orthogonal to the signal regions,  
with identified electrons or muons in the final state  and with the same requirements on 
the jet $\ptjet$, subleading jet vetoes, and $\met$, are employed to determine 
the normalization of the electroweak backgrounds from data.
This  reduces significantly the relatively large  theoretical and experimental 
systematic uncertainties associated to purely MC-based predictions. 
The muon control sample is used to normalize  the $\Wmnj$+jets, $\Znn$+jets, and $\Zmm$+jets 
 MC predictions. Similarly, the electron sample is employed to determine the 
normalization of the $\Wenj$+jets, $\Zee$+jets, and 
$\Wtauj$+jets MC  predictions.

\subsection{Multi-jets background}

 The multi-jets background  with large $\met$
originates mainly from the misreconstruction of the energy of a  jet 
in the calorimeters, which goes below the required 
threshold, resulting in  a monojet signature.
In such events, the $\met$ direction will generally be aligned 
with the second-leading jet in the event. 
To estimate this background, jets-enriched data
control samples  are defined using the 
$\LowPt$, $\HighPt$, and $\veryHighPt$ selections without the veto on the second-leading 
jet $\pt$ and requiring $\Delta \phi (jet2 - \met) < 0.5$.  Events with 
more than two jets with $\pt$ above 30~GeV   
are excluded. The measured $\pt$ distribution of the 
second-leading jet in the multi-jets enriched control samples ($\Delta \phi (jet2 - \met) < 0.5$) are used 
to estimate the multi-jets background in the analyses. 
The number of multi-jets background events is obtained from a linear extrapolation 
below the threshold  of $\pt < 30$~GeV ($\pt < 60$~GeV) for the $\LowPt$ ($\HighPt$ and $\veryHighPt$) region.

\subsection{Non-collision background}

The contribution of non-collision backgrounds to the selected  monojet samples 
from cosmics rays,   
overlaps between  background
events and  genuine proton-proton collisions, and from beam-halo muons 
are estimated using  events registered in empty  and 
unpaired proton bunches in the collider that fulfill the event 
selection criteria, and a beam-halo tagger.
The latter combines information
from the muon chambers and the timing of calorimeter clusters.

The expected background predictions are summarized in Table 1 for the $\LowPt$, $\HighPt$, and $\veryHighPt$ selections. 
Good agreement is observed between the data and the SM predictions in all cases.

\begin{table}[htbp]
\begin{center}
\begin{tiny}
\renewcommand{\baselinestretch}{1.2}
\begin{tabular}{c c c c} \hline\hline
\multicolumn{4}{c}{\tiny{Background  Predictions $\pm$ (stat.) $\pm$ (syst.)}} \\ 
                    & $\LowPt$ Selection &  $\HighPt$ Selection & $\veryHighPt$ selection\\  \hline
$\Znn$+jets                            & $7700\pm90\pm400$    & $610\pm27\pm47$      & $124\pm12\pm15$\\
$\Wtauj$+jets                          & $3300\pm90\pm220$    & $180\pm16\pm22$      & $36\pm7\pm8$\\
$\Wenj$+jets                           & $1370\pm60\pm90$     & $68\pm10\pm8$        & $8\pm1\pm2$\\
$\Wmnj$+jets                           & $1890\pm70\pm100$     & $113\pm14\pm9$       & $18\pm4\pm2$\\
Multi-jets                         & $360\pm20\pm290$     & $30\pm6\pm11$        & $3\pm2\pm2$\\
$\Ztt$+jets                            & $59\pm3\pm4$         & $2.0\pm0.6\pm0.2$    & -\\
$\Zmm$+jets                            & $45\pm3\pm2$         & $2.0\pm0.6\pm0.1$    & -\\
$t\bar{t}$                             & $17\pm1\pm3$         & $1.7\pm0.3\pm0.3$    & -\\
$\gamma$+jet                           & -                & -                    & -\\
$\Zee$+jets                            & -                    & -                    & -\\
Non-collision Background               & $370\pm40\pm170$     & $8.0\pm3.3\pm4.1$    & $4.0\pm3.2\pm2.1$ \\ \hline
Total Background                       & $15100\pm170\pm680$  & $1010\pm37\pm65$     & $193\pm15\pm20$\\
Events in Data  (1.00 fb${}^{-1}$)        & $15740$        & $965$           & $167$\\\hline
 \end{tabular}
\end{tiny}
\end{center}
\renewcommand{\baselinestretch}{1}
\label{tab:Results}
\caption{
Number of observed events and predicted  background events, including statistical and
systematic uncertainties. 
The statistical uncertainties are due to limited MC statistics. 
The dominant systematic uncertainties come from the limited statistics in the data control regions.
The systematic uncertainties on $\Wmnj$+jets, $\Zmm$+jets, and $\Znn$+jets predictions 
are fully correlated.  Similarly, the systematic uncertainties on $\Wenj$+jets, $\Wtauj$+jets, and $\Ztt$+jets are 
fully correlated. 
}
\end{table}

\section{Results}
\label{sec:results}

\begin{figure}[h]
\begin{center}
\resizebox{0.78\columnwidth}{!}{%
\includegraphics{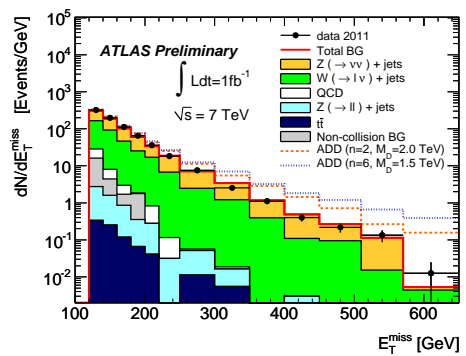}}
\end{center}
\caption{\small 
Measured $\met$ distribution (black dots) 
in the $\LowPt$ region compared to 
the predictions for SM backgrounds (histograms). Only statistical uncertainties are considered.
For illustrative purposes, the impact of two different ADD scenarios is included.   
}
\label{fig:low}
\end{figure}

Figure~\ref{fig:low} shows the measured 
$\met$ distribution for the $\LowPt$
selection compared to the background predictions. For illustrative purposes, the 
Figure indicates the impact of two different ADD scenarios. 

\subsection{Model-independent limits on $\sigma \times$ Acceptance}

The agreement between the data and the SM predictions for the total number of events  
in the different analyses is translated into model-independent 95$\%$ confidence level (CL) upper limits 
on the cross section times acceptance. The 
$CL_s$  modified frequentist approach is used and the upper limits are 
computed considering the  uncertainties on the background predictions (see Table~1) 
and a 4.5$\%$ uncertainty on the quoted integrated luminosity. The resulting 
95$\%$ CL limits on cross section times acceptance for the $\LowPt$, $\HighPt$, and $\veryHighPt$ selections are 1.7~pb, 0.11~pb, and 0.035~pb, respectively.

The data are used to set new improved 95$\%$ CL upper limits on the parameters of the ADD LED model.
MC simulated samples for the ADD LED model with 
different number of extra dimensions varying from 2 to 6 are generated using the PYTHIA program with
CTEQ6.6 PDFs, and renormalization and factorization scales set to 
$\frac{1}{2} M_{G}^2 + p_{T}^2$,
where $M_G$ is the graviton mass and $p_T$ 
denotes the transverse momentum of the recoiling parton.
As already pointed out in~\cite{plb}, 
the effective theory used to compute the ADD cross sections  remains valid only if the scales involved 
in the hard interaction are smaller than $M_D$. Otherwise, the predictions strongly depend  on the 
unknown ultraviolet behavior of the theory. 

\begin{figure}[h]
\begin{center}
\resizebox{0.78\columnwidth}{!}{%
\includegraphics{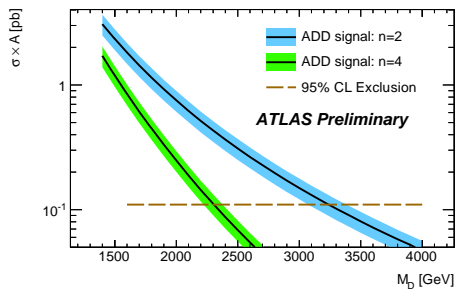}}
\end{center}
\caption{
Cross section times acceptance curves as a function of $M_D$ 
predicted by the effective theory for 2 and 4  
extra dimensions. 
The bands surrounding the curves reflect the systematic uncertainties on the predicted signal yields.
The model-independent cross section times acceptance limit is shown as a dashed line for the $\HighPt$ region.
}
\label{fig:limit1}
\end{figure}

Figure~\ref{fig:limit1} shows the predicted ADD cross section times acceptance in the $\HighPt$ region 
as a function of $M_D$ for 2 and 4 extra dimensions, where the band reflects the  total 
uncertainty on the signal. For illustration purposes, the model-independent limit of 0.11~pb is included. 

The simulation indicates that
the detector effects reduce the expected signal yields, as determined at the particle
level, by a factor $0.83 \pm 0.01$,  
approximately independent of $M_D$ and $n$.
This results into  
95$\%$ CL cross section upper limits of 2.02~pb, 0.13~pb, and 0.045~pb for the $\LowPt$, $\HighPt$, and $\veryHighPt$ 
regions, respectively.

\subsection{Lower limits on $M_D$}

\begin{figure}[h]
\begin{center}
\resizebox{0.78\columnwidth}{!}{%
\includegraphics{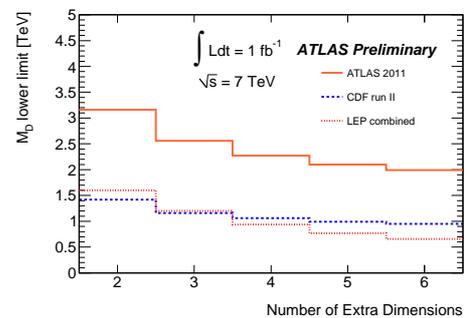}}
\end{center}
\caption{
The 95\% CL observed lower limits on $M_D$ for different numbers of extra dimensions for ATLAS, 
compared with previous results.
}
\label{fig:limit2}
\end{figure}

New improved 95$\%$ CL lower limits are set on the value of $M_{D}$ as a function of 
the number of extra dimensions considered in the ADD LED model. The $CL_s$ approach is used, 
including statistical and systematic uncertainties. The $\HighPt$ selection is used for the final results. It 
provides better expected limits than the ones obtained in the 
$\LowPt$ region, and the results are comparable with those  
for the $\veryHighPt$ region (mainly due to the rapid decrease of the signal cross section with 
increasing jet $\ptjet$ and $\met$) but  with a reduced sensitivity to the ultraviolet behavior of the theory~\cite{atlas_eps}.
Figure~\ref{fig:limit2} presents
the observed 95$\%$ CL lower limits on $M_D$ as a function of the number of extra dimensions varying from 2 to 6, as 
determined by the $\HighPt$ selection.
This analysis imposes 95$\%$ CL 
upper limits for the scale $M_D$  that vary between 3.2~TeV for $n=2$ and 2.3~TeV for $n=4$ to 
2.0~TeV for $n=6$, significantly extending previous results.  

%


\section{Summary and conclusions}
\label{sec:sum}
In this note we have presented preliminary results on the search for 
new phenomena in an event sample with monojets and large missing transverse 
momentum in the final state, based on 1.00~fb${}^{-1}$ of data collected by 
the ATLAS experiment in 2011. Good agreement is observed between the 
data and the Standard Model predictions. The results are translated into 
model-independent 95$\%$ confidence level upper limits on fiducial cross sections that vary 
between 2.02 pb and 0.045 pb for the $\LowPt$ and $\veryHighPt$ analysis, respectively.
The results are also interpreted in terms of the ADD LED scenario for which
$M_D$ values between 3.2~TeV and 2.0~TeV are excluded at the 95$\%$ confidence level
for a number of extra dimensions varying from 2 to 6, respectively. These results
significantly improve previous limits on models with Large Extra Dimensions.



\end{document}